\documentstyle[11pt,newpasp,twoside,epsf]{article}
\begin{document}
\title{Joint VLTI/VLBA observations of Mira stars}
\author{Markus Wittkowski}
\affil{European Southern Observatory, Garching bei M\"unchen, Germany}
\author{David A. Boboltz}
\affil{U.S. Naval Observatory, Washington, DC, USA}

\begin{abstract}
We present preliminary results on a recently started project to perform
coordinated observations of Mira stars using near-infrared and radio
long-baseline interferometry. We concentrate on recent observations
of the Mira star S\,Ori.
Observations with the ESO Very Large Telescope Interferometer (VLTI)
were performed to measure the near-infrared diameter of the stellar surface.
Concurrent VLBA observations of SiO maser emission towards this Mira star
were performed to probe the structure and dynamics of the circumstellar
atmosphere. Our near-infrared measurements suggest a diameter change 
consistent with the photometric period. The SiO maser emission
appears to be at a distance of $\sim$\,2 stellar radii 
( $\Theta_{\rm UD} \sim$\,10.7\,mas, 
$\Theta_{\rm SiO} \sim$\,20\,mas, both at stellar phase $\sim$\,0.7),
and to show a clumpy distribution within a ring-like structure.
\end{abstract}

\section{Introduction}
\begin{figure}[t]
\vspace*{-5cm}
\plotone{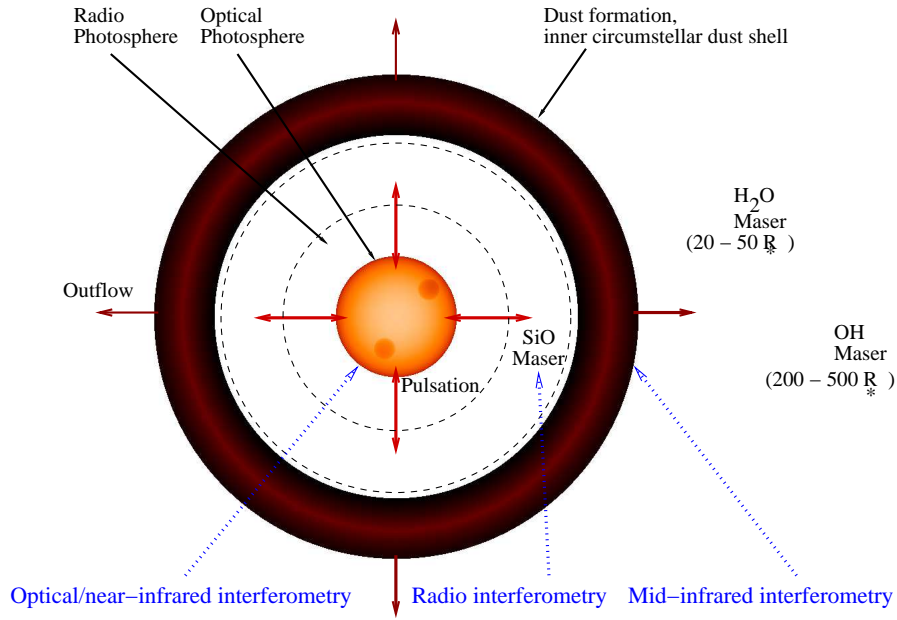}
\vspace*{-5.5cm}
\caption{Multi-wavelength view of a Mira star and its circumstellar envelope.
Near-infrared interferometry can probe the stellar surface, 
radio interferometry the SiO maser emission at typically $\sim$\,2--3 
stellar radii and mid-infrared interferometry
the dust formation shell at distances of typically 4--10 stellar radii. 
Radio interferometry can also probe the circumstellar envelope at distances
of about 20--500 stellar radii by observing H$_2$O and OH maser emission.}
\label{fig:schema}
\end{figure}
The evolution of late-type stars along the asymptotic giant branch
(AGB) is accompanied by significant mass loss to the circumstellar
environment with mass-loss rates of up to $10^{-7}-10^{-4}$\,M$_\odot$/year
(see e.g., Jura \& Kleinmann 1990, Kemball \& Diamond 1997). 
AGB stars with masses of the order of 1\,M$_\odot$ become unstable to
large-amplitude radial pulsations and become Mira variable stars.
The nature of the mass-loss process of AGB stars, and especially
its connection with the pulsation mechanism in the case of Mira variable stars, 
is not well understood.
Joint multi-wavelength studies of the stellar surface and the
circumstellar environment at different distances from the stellar surface
promise to lead to a better understanding of the nature of the 
mass-loss process, 
and thus of the evolution of AGB stars and of the chemical 
enrichment process of galaxies.
Figure~1 shows a scheme of the multi-wavelength view of a Mira star and
its circumstellar environment.
The conditions on the stellar surface can in principle be studied 
by means of optical/near-infrared long-baseline interferometry.
This technique provides information regarding the stellar
diameter, effective temperature, and center-to-limb intensity variations.
With multiple epochs, the variability of
these parameters as a function of time/stellar pulsation cycle can 
be studied as well. Interferometry at a more advanced stage can likely
probe horizontal temperature inhomogeneities of the stellar surfaces 
as well (e.g. Wittkowski et al. 2002).
Recent results on Mira stars based on near-infrared interferometry 
are described by e.g. Thompson et al. (2002a/b).
The ESO Very Large Telescope Interferometer (VLTI) provides excellent 
capabilities to observe AGB stars at near-infrared and 
mid-infrared wavelengths. It is currently in the phase of commissioning.
About 20 Mira stars have already been observed with the near-infrared
commissioning instrument VINCI (Richichi \& Wittkowski 2003).
The structure and dynamics of the circumstellar atmosphere of Mira stars
at typically 2--3 stellar radii can be probed by mapping the 
SiO maser emission towards these stars using very long baseline 
interferometry (VLBI) at radio wavelengths
(e.g., Kemball \& Diamond 1997, Boboltz et al. 1997,
Boboltz \& Marvel 2000, Boboltz 2003). The dust formation zone at distances of 
typically 4--10 stellar radii can be probed by mid-infrared interferometry
(see, e.g. Danchi et al. 1994, Townes 2003). The distance of the 
SiO maser emission relative to the dust zone has been 
studied by Greenhill et al. (1995).

Here, we report on first preliminary results of a recently
started project to combine observations of the SiO maser
emission towards Mira stars using the VLA/VLBA and observations
of the near-infrared stellar disk using the VLTI, focusing
on the Mira star S\,Ori.
\section{Characteristics of S\,Ori}
S\,Ori is a relatively cool Mira star with spectral type M\,6.5\,e -- M\,9.5\,e
and relatively large visual amplitude of $V$\,$\sim$\, 7--13. 
Merchan Benitez \& Jurado Vargas (2000) report on a marked decrease of the
stellar variability period, which might be caused by a recent 
helium-shell flash.
The variability period is currently $\sim$\,440 days. The distance to S\,Ori
was estimated by van Belle (1996) to 422\,$\pm$\,37\,pc (based on Wyatt\& Cahn 1993 
and Young 1995).
There is one previous diameter measurement of S\,Ori by van Belle et al. (1996),
who derived a uniform disk (UD) diameter of 10.54\,$\pm$\,0.68\,pc at
stellar pulsation phase 0.56. SiO and OH maser emission towards this
star has been reported.
An interferometric map of this emission
has so far not been obtained.

We have chosen the object S\,Ori for our study because it is known to show 
a fairly strong (10-20 Jy) SiO maser emission, it has a size and K magnitude
which is feasible for observations with the VLTI and 
the 40\,cm test siderostats,
and it was available for both facilities, VLTI and VLBA.
\section{VLTI and VLBA measurements}
\begin{figure}[t]
\plottwo{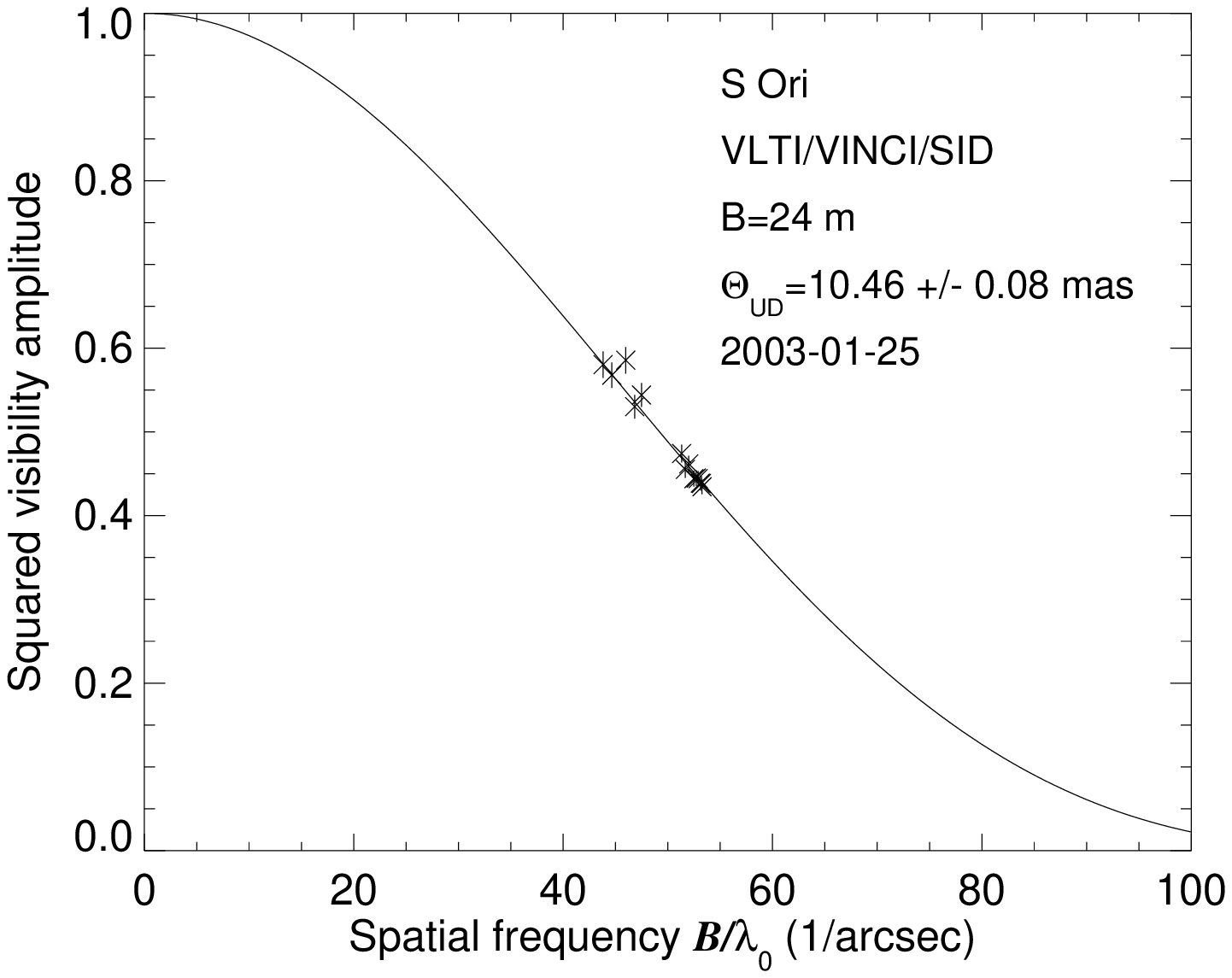}{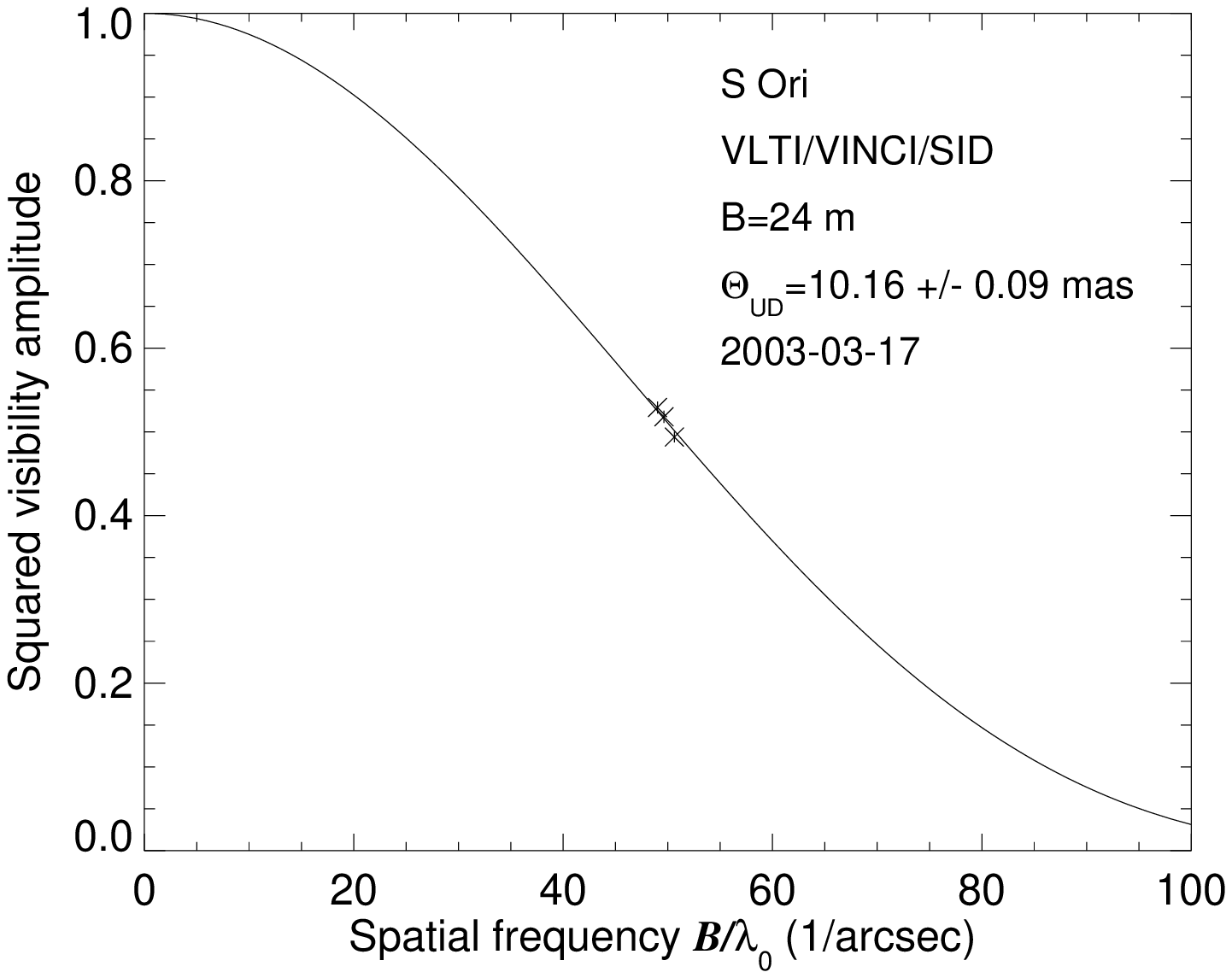}
\caption{Two examples of our VLTI visibility curves for S\,Ori}.
\end{figure}
\begin{figure}[t]
\textwidth=0.8\textwidth
\plotone{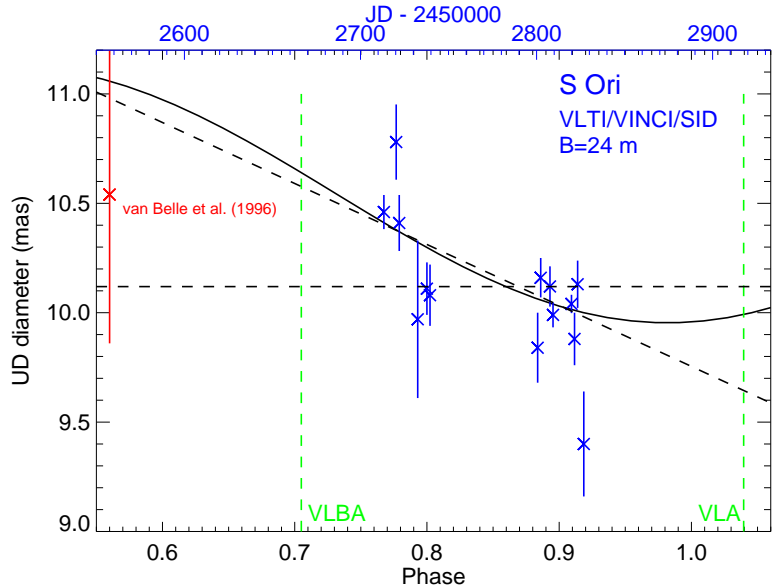}
\textwidth=1.25\textwidth
\caption{UD diameter of S\,Ori as a function of visual variability phase}.
\end{figure}
$K$-band interferometric data on S\,Ori were obtained on 14 days between January
and March 2003, corresponding to stellar phases $\sim$\,0.77 -- 0.92, 
using the VLTI with the commissioning instrument VINCI, the 40\,cm
test siderostats, and a baseline of 24\,m.
These observations have been coordinated with
VLBA observations of the SiO maser emission towards our source, which were
obtained on Dec. 29, 2002, at stellar phase $\sim$\,0.7. These observations
were carried out for two maser transitions, namely $v=2, J=1-0$, 42.8\,GHz,
and $v=1, J=1-0$, 43.1\,GHz.

Figure~2 shows two examples of our preliminary VLTI visibility curves for 
S\,Ori, for dates January 25 and March 17. They show that a uniform disk
model fits our data. There is no obvious indication of deviations from
a uniform disk model at these spatial frequencies which might in principle
be caused by contributions from a circumstellar envelope. 
Figure~3 shows our obtained near-infrared $K$-band UD diameters as a 
function of date and stellar 
variability phase. Indicated are also the UD diameter measurement by
van Belle (1996), and the date of our coordinated VLBA measurements.  
Our UD diameter values suggest a decrease of the UD diameter over our
time period, which is consistent with the stellar phase. The dotted and
solid lines indicate best fits of a linear function and a sine function,
respectively. 

The obtained uniform disk diameters will be transformed into more physically
meaningful Rosseland-mean diameters using stellar atmosphere models
(cf. Hofmann et al. 1998, Wittkowski et al 2003).

\begin{figure}[t]
\plottwo{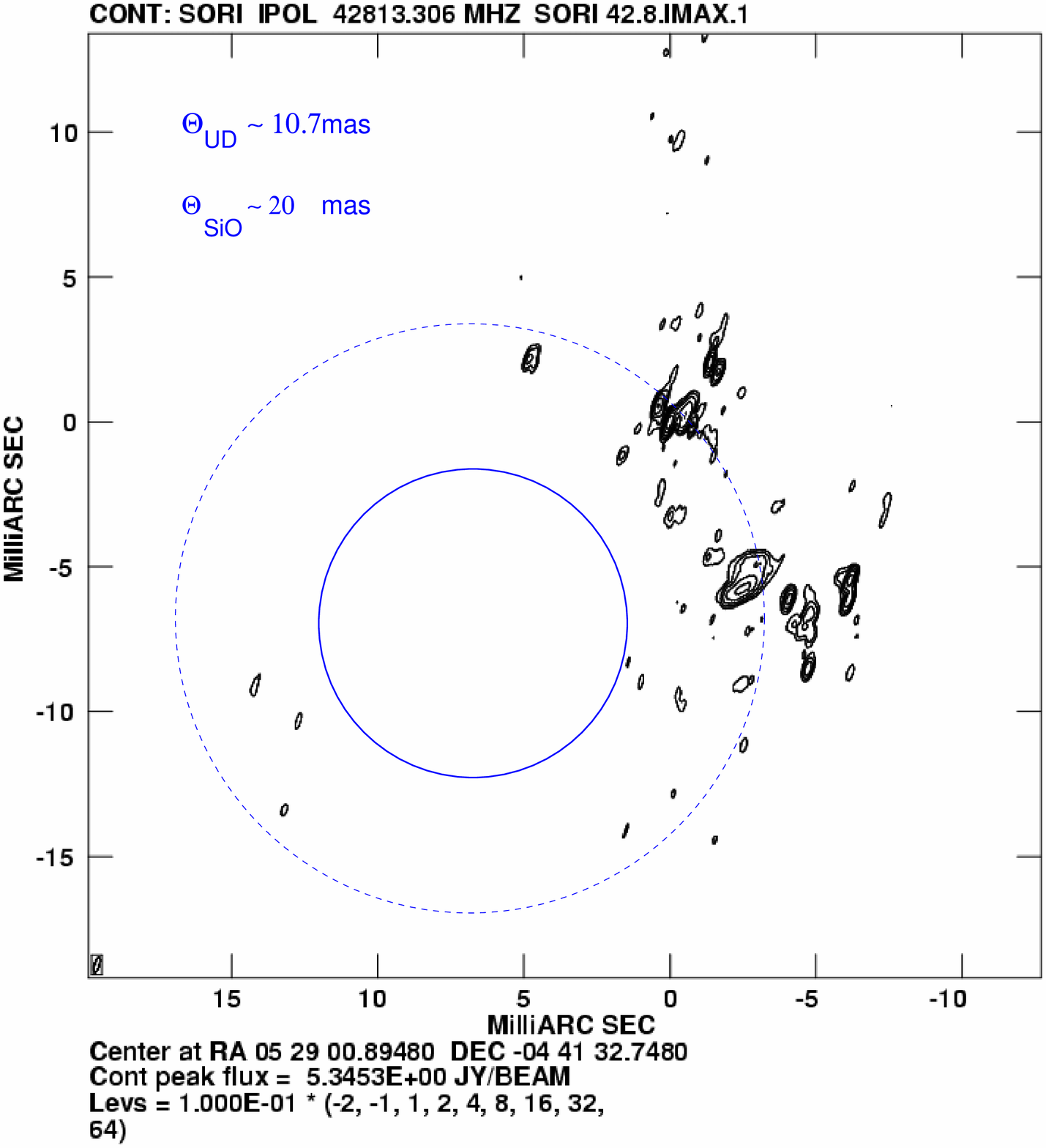}{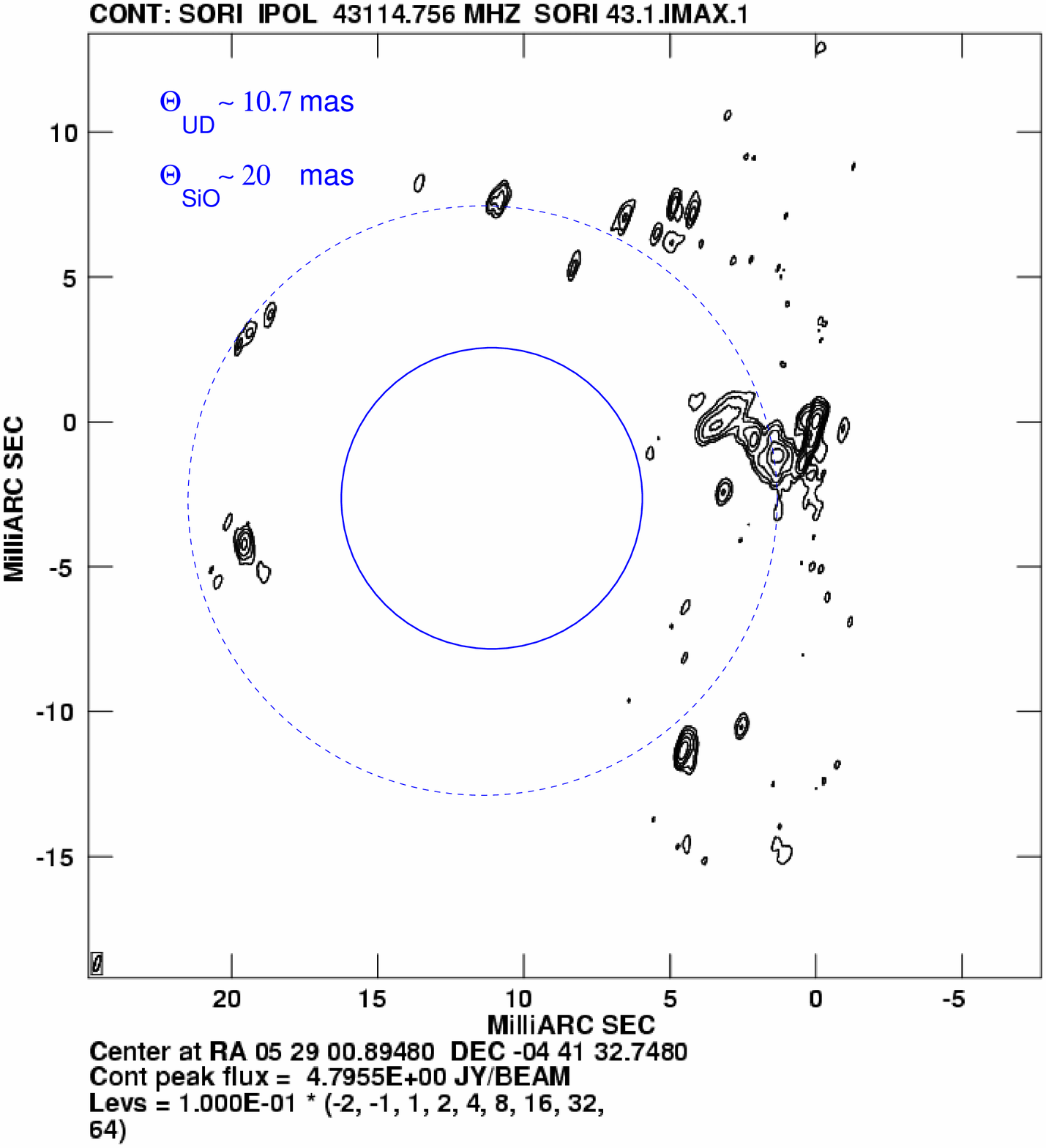}
\caption{SiO maser emission towards S\,Ori (Left: $v=2, J=1-0$; 
42.8\,GHz transition. Right: $v=1, J=1-0$; 43.1\,GHz transition).
Shown are also the near-infrared stellar diameter as obtained 
by our VLTI measurements, as well as an approximate distance 
of the SiO maser emission regions.}
\end{figure}

Figure~5 shows our obtained VLBA maps of the SiO maser emissions towards
S\,Ori. Both transitions show a clumpy distribution within a ring-like
structure. Indicated are also our obtained near-infrared $K$-band UD
diameter at the phase of the VLBA map and an approximate mean distance
of the SiO maser emission spots to the center of the star. The relative
positions of the near-infrared diameter and the SiO maser emission
are arbitrary. The emission from both SiO maser 
transmissions appear to be at a distance of $\sim$\,2 stellar radii
($\Theta_{\rm UD}$\,$\sim$\,10.7\,mas\,$\sim$\,4.5\,AU;
$\Theta_{\rm SiO}$\,$\sim$\,20\,mas\,$\sim$\,8.4\,AU).
The emissions from the two SiO maser transmissions (42.8\,GHz and 43.1\,GHz)
appear to have a separation of $<$\,0.4\,AU, which is consistent with
predictions by Gray \& Humphreys (2000). 

Future work on this project aims at obtaining near-infrared visibility values
at more stellar phases and spatial frequencies to confirm the UD diameter change
with stellar variability phase and to constrain the intensity profile, respectively.
Multi-epoch VLBA maps would allow us to correlate the near-infrared 
diameter change with changes of the distribution of the SiO maser emission.
Mid-infrared interferometric measurements could in addition be used to 
probe the distribution and distance of the dust formation shell.

\vspace*{-1.0cm}%

{\flushleft\small The VLTI data were taken in the framework of our 
shared risk program in ESO observing period 70, and are public data 
released from the ESO VLTI.}


\begin{references}
\reference Boboltz, Diamond, Kemball, 1997, ApJ 487, L147
\reference Boboltz \& Marvel, 2000, ApJ 545, L149
\reference Boboltz, 2003, this volume
\reference Danchi et al., 1994, AJ 107, 1469
\reference Greenhill et al., 1995, ApJ 449, 365
\reference Hofmann, Scholz, Wood, A\&A 339, 846
\reference Jura \& Kleinmann, 1990, ApJS 73, 769
\reference Kemball \& Diamond, 1997, ApJ 481, L111
\reference Merch\'an Ben\'itez P., Jurado Vargas M., 2002, A\&A 386, 244
\reference Thompson, Creech-Eakman, Akeson, 2002a, ApJ. 570, 373-378
\reference Thompson, Creech-Eakman, van Belle, 2002b,  ApJ 577, 447
\reference Townes, 2003, in Reviews of Modern Astronomy 16, E. Schilicke ed., p. 1
\reference Richichi, Wittkowski, 2002, Ap\&SS, in press
\reference van Belle, Dyck, Benson, Lacasse, 1996, AJ 112, 2147 
\reference Wittkowski, Sch\"oller, Hubrig, Posselt, von der L\"uhe, 2002, AN 323, 241
\reference Wittkowski, Aufdenberg, Kervella, 2003, submitted to A\&A
\reference Wyatt, Cahn, 1983, ApJ 275, 225
\reference Young, 1995, ApJ 445, 872
\end{references}
\end{document}